\documentclass[reprint,amsmath,amssymb,aps,prl]{revtex4-1}
\pdfoutput=1
\usepackage{amsmath}
\usepackage{amssymb}
\usepackage{graphicx}

\usepackage{dcolumn}
\usepackage{bm}
\usepackage{hyperref}

\begin{document}

\title{Mutual inactivation of Notch and Delta permits a simple mechanism for lateral inhibition patterning}

\author{Amit Lakhanpal}
 \email{amitl@caltech.edu}
\author{David Sprinzak}
 \email{davidsp@caltech.edu}
\author{Michael B. Elowitz}
 \email{melowitz@caltech.edu}
\affiliation{
 Departments of Biology and Applied Physics, California Institute of Technology
}
\affiliation{
 Howard Hughes Medical Institute
}

\date{\today}

\begin{abstract}
Lateral inhibition patterns mediated by the Notch-Delta signaling system occur in diverse developmental contexts.  These systems are based on an intercellular feedback loop in which Notch activation leads to down-regulation of Delta.  However, even in relatively well-characterized systems, the pathway leading from Notch activation to Delta repression often remains elusive.  Recent work has shown that \emph{cis}-interactions between Notch and Delta lead to mutual inactivation of both proteins.  Here we show that this type of \emph{cis}-interaction enables a simpler and more direct mechanism for lateral inhibition feedback than those proposed previously.  In this mechanism, Notch signaling directly up-regulates Notch expression, thereby inactivating Delta through the mutual inactivation of Notch and Delta proteins.  This mechanism, which we term Simplest Lateral Inhibition by Mutual Inactivation (SLIMI), can implement patterning without requiring any additional genes or regulatory interactions.  Moreover, the key interaction of Notch expression in response to Notch signaling has been observed in some systems.  Stability analysis and simulation of SLIMI mathematical models show that this lateral inhibition circuit is capable of pattern formation across a broad range of parameter values.  These results provide a simple and plausible explanation for lateral inhibition pattern formation during development.
\end{abstract}

\maketitle

\section{Introduction}
\begin{figure}[htbp]
\centering
\includegraphics{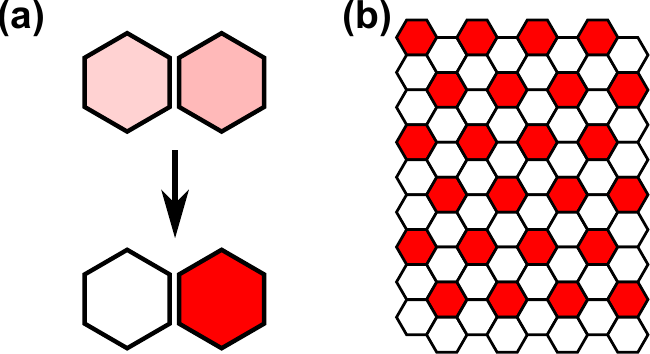}
\caption{(a) Two initially near-equivalent cells in direct contact eventually reach very different final states, characteristic of `fine-grained' patterning. (b) An ideal lateral inhibition pattern in a two-dimensional field of cells.}
\label{fig:LI_cartoons}
\end{figure}

Multicellular development often involves transitions from initially near-homogeneous tissues to `fine-grained' patterns involving sharp distinctions between neighboring cells.  One such pattern is ``lateral inhibition'' (LI), characterized by alternating patterns of `on' and `off' states such as those diagrammed in Fig. \ref{fig:LI_cartoons}.  This phenomenon is pervasive, arising in situations as diverse as butterfly wing coloration \cite{reed_2004}, neuroectoderm specification \cite{campos-ortega_1994}, ciliated cell specification \cite{marnellos_2000}, and sensory organ precursors \cite{heitzler_1991}.

\begin{figure}[htbp]
\centering
\includegraphics[width=0.7\columnwidth]{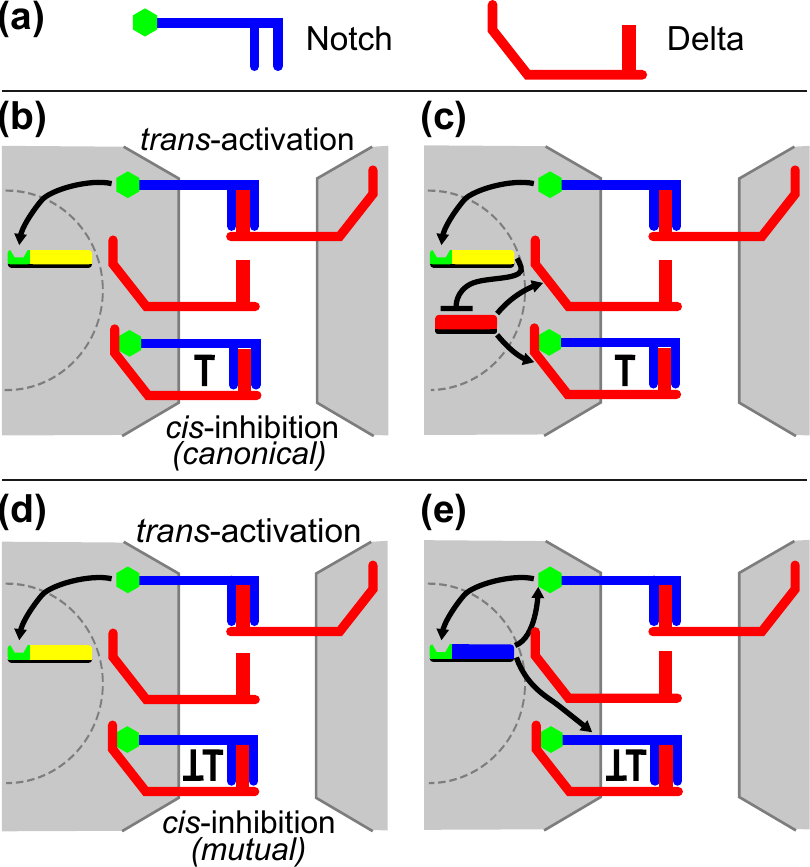}
\caption{(a) Notch (blue) and Delta (red). (b) Notch and Delta interact in \emph{trans} (on neighboring cell surfaces) to send the intracellular Signal domain (green) of Notch to the nucleus.  Delta also inhibits Notch in \emph{cis} (on the same cell surface). (c) A canonical lateral inhibition feedback in which Notch signaling induces expression of an intermediate (yellow) that represses Delta expression.  Although not explicitly drawn in the figure, the same feedback operates in all cells.  (d) The \emph{trans} and \emph{cis} interactions of Notch and Delta, with a \emph{mutual} inactivation of the receptor and ligand in \emph{cis}.  (e)  A surprisingly simple lateral inhibition feedback network in which Notch signaling induces Notch expression, which directly inactivates Delta by the mutual \emph{cis}-inactivation mechanism.}
\label{fig:notch_delta}
\end{figure}

LI patterning in these and other contexts is mediated by signaling through the Notch-Delta system.  The Notch-Delta system (reviewed in \cite{weinmaster_2008,weinmaster_kopan_2006, bray_2006, artavanis-tsakonas_1999}) consists of the Notch receptor family and its Delta-family ligands (blue and red molecules, respectively, in Fig. \ref{fig:notch_delta}), along with numerous participants in the signaling mechanism.  Delta interacts with Notch in two modes (Fig. \ref{fig:notch_delta}b):  activating Notch signaling in neighboring cells (\emph{trans}-activation) while inhibiting Notch signaling in the same cell (\emph{cis}-inhibition).  LI patterning can occur when Notch signaling downregulates Delta levels.  This downregulation is usually assumed to be mediated by a transcriptional repressor, although it could also be implemented through post-transcriptional mechanisms (Fig. \ref{fig:notch_delta}c).  Under certain conditions \cite{lewis_1996, plahte_2007} a high level of Delta in one cell will drive all of its neighbors to low levels of Delta expression.  Conversely, a cell whose neighbors are all Delta-poor will eventually express Delta at a high level.  This generates the lateral inhibition pattern of Fig. \ref{fig:LI_cartoons}b with high-Delta cells (red) surrounded by low-Delta cells (white).

The feedback pathway inhibiting Delta expression in response to Notch activation may be known in certain cases \cite{lecourtois_1995}.  However, in many contexts it remains unclear what components, if any, play this role \cite{greenwald_1998}.  On the other hand, in some natural systems such as vein patterning \cite{huppert_1997} and lateral inhibition \cite{wilkinson_1994}Notch activation is known to induce Notch expression.  Further, a recent quantitative study of the Notch signaling response function uncovered evidence for \emph{mutual} \emph{cis}-inhibition \cite{sprinzak_2010}.  Not only does Delta inactivate Notch signaling in \emph{cis} as drawn in Fig. \ref{fig:notch_delta}b, but Notch also reciprocally inactivates Delta as drawn in Fig. \ref{fig:notch_delta}d.

Here we report that the mutual inactivation model of \emph{cis}-inhibition admits the possibility of a remarkably simple mechanism for achieving lateral inhibition.  Notch signaling upregulation of Notch receptor expression, combined with the mutual inactivation mechanism, directly downregulates Delta levels in response to Notch signaling (Fig. \ref{fig:notch_delta}e).  Mathematical analysis of this feedback circuit shows that it can generate the LI pattern and provides some advantages compared to the canonical architecture.  We describe this as the Simplest Lateral Inhibition by Mutual Inactivation (SLIMI) model.

\section{Results}
\begin{figure}[htbp]
\centering
\includegraphics[width=0.7\columnwidth]{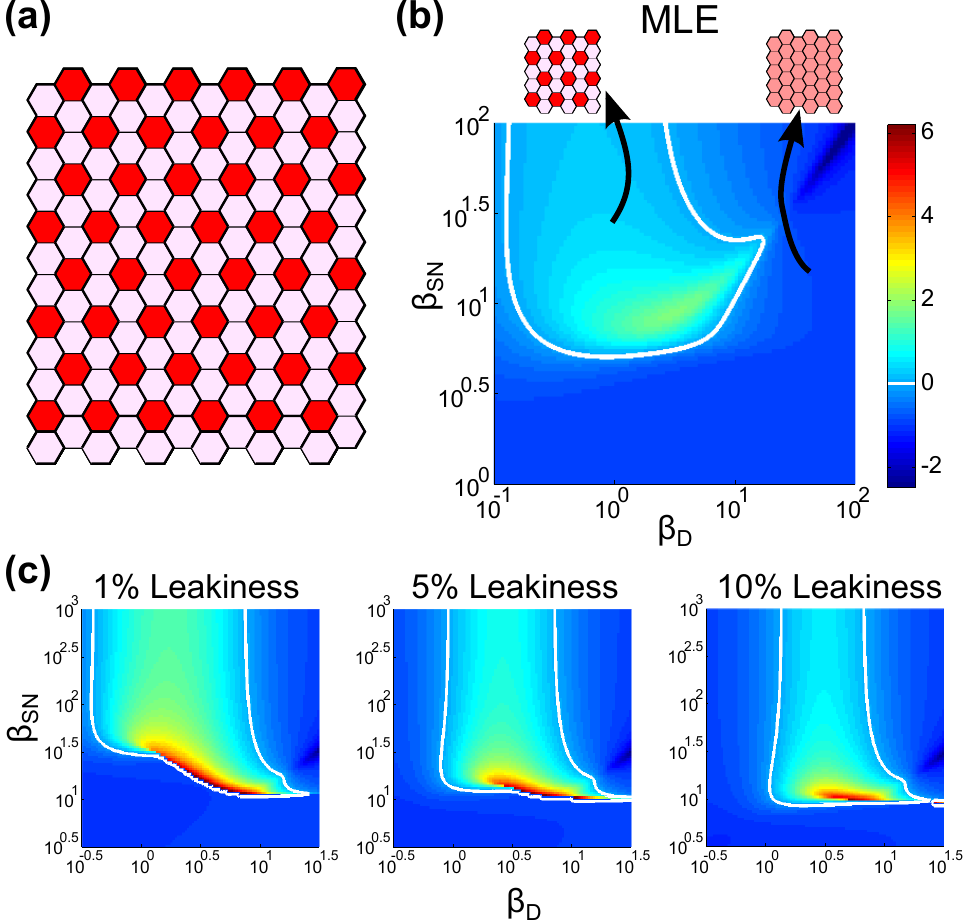}
\caption{(a) Outcome of a numerical simulation of the SLIMI mechanism colored as in Fig. \ref{fig:LI_cartoons}b. (b) Maximum Lyapunov Exponent (MLE) computed across a range of Delta production rates $\beta_\text{D}$ and maximal Notch production rates $\beta_{\text{SN}}$, with other parameters fixed including $n=1$.  The region within the contour is above zero, indicating instability of the homogeneous steady state and thus the potential to generate the LI pattern.  (c) MLE computed at $n=2$ with increasing levels of `leakiness' in Notch expression, leading to some shrinkage of the patterning-permissive parameter range.}
\label{fig:SLIMI}
\end{figure}

In order to analyze the SLIMI circuit, we assume an ideal two-dimensional lattice of hexagonal cells, each containing Notch ($\text{N}_{i}$) and Delta ($\text{D}_{i}$) interacting in \emph{trans} to generate Signal ($\text{S}_{i}$) that induces Notch expression, and in \emph{cis} with mutual inactivation.  Correspondingly, we consider the following reactions:
\begin{eqnarray}
&\text{N}_i + \text{D}_j \rightleftharpoons [\text{N}_i \text{D}_j] \rightarrow \text{S}_i &\phantom{aaa}\text{  \emph{trans}-activation} \label{eqn:trans-activation}\\
&\text{N}_i + \text{D}_i \rightleftharpoons [\text{N}_i \text{D}_i] \rightarrow \varnothing &\phantom{aaa}\text{  \emph{cis}-inactivation} \label{eqn:cis-inhibition}\\
&\text{S}_i \rightarrow \text{N}_i &\phantom{aaa}\text{  Notch induction} \label{eqn:notch_induction}
\end{eqnarray}
Reaction \ref{eqn:trans-activation} refers to \emph{trans}--activation in cell $i$ by ligand on neighboring cells $j$, with Notch-Delta \emph{trans} association (dissociation) rate $k_{D}^{+}$ $(k_{D}^{-})$ and signal release rate $k_{S}$.  Reaction \ref{eqn:cis-inhibition} refers to \emph{cis}--inhibition, with Notch-Delta \emph{cis} association (dissociation) rate $k_{C}^{+}$ $(k_{C}^{-})$ and mutual inactivation rate $k_{E}$.  Reaction \ref{eqn:notch_induction} refers to Signal activation of Notch expression, which we parametrize as a contribution to the rate of Notch production in the form of an increasing Hill function ($\beta_{\text{SN}}\frac{\text{S}_{i}^{n}}{K_{\text{SN}}^{n}+\text{S}_{i}^{n}}$).  Allowing for ``leakiness'' in Notch production (non-zero production rate $\beta_{\text{N}}$ in the absence of inducer), constant constitutive production of Delta ($\beta_\text{D}$), linear degradation of each component($-\gamma\text{N}_{i}$ and $-\gamma\text{D}_{i}$), and taking the quasi-steady-state approximation on the receptor-ligand complexes and the Signal molecule, these reactions translate to the following set of ordinary differential equations:
\begin{eqnarray}
\dot{\text{N}}_i = &&\beta_\text{N} + \beta_{\text{SN}}\frac{\text{N}_{i}^{n}\left\langle\text{D}_{j}\right\rangle_{i}^{n}}{K_{\text{SN}}^{n}+\text{N}_{i}^{n}\left\langle\text{D}_{j}\right\rangle_{i}^{n}} - \gamma\text{N}_{i} \nonumber \\ &&-\text{N}_{i}\frac{\left\langle\text{D}_{j}\right\rangle_{i}}{k_{t}}-\text{N}_{i}\frac{\text{D}_{i}}{k_{c}} \\
\dot{\text{D}}_i =&& \beta_{\text{D}} - \gamma\text{D}_{i} - \left\langle\text{N}_{j}\right\rangle_{i}\frac{\text{D}_{i}}{k_{t}}-\text{N}_{i}\frac{\text{D}_{i}}{k_{c}}
\end{eqnarray}
Here we have employed the notation $\left\langle\cdot_{j}\right\rangle_{i}$ to denote the average of the enclosed quantity among the neighbors $j$ of cell $i$.  We define the parameters $k_{t}^{-1}\equiv\frac{k_{\text{D}}^{+}k_{\text{S}}}{k_{\text{D}}^{-}+k_{\text{S}}}$ and  $k_{c}^{-1}\equiv\frac{k_{\text{C}}^{+}k_{\text{E}}}{k_{\text{C}}^{-}+k_{\text{E}}}$ to denote the strengths of the \emph{trans} and \emph{cis} interactions, respectively.  Numerical simulation of these equations proves that they are capable of generating the LI pattern from a slightly (and randomly) heterogeneous field of cells (Fig. \ref{fig:SLIMI}a).

In order to more generally determine conditions under which this system of coupled, non-linear differential equations can generate the LI pattern, we performed a linear stability analysis about the system's homogeneous steady state (hss).  If the hss is stable (unstable) to small perturbations, it follows that the LI pattern is inaccessible (accessible) from an initial condition near the hss.  This analysis required computing and diagonalizing the Jacobian evaluated at the hss, and the system's stability there was determined by the sign of the maximal eigenvalue (known as the Maximum Lyapunov Exponent --- MLE).  Where the MLE is positive the hss is unstable, and the LI pattern is accessible.

Computing the MLE requires knowledge of component leves at the hss, defined as the solutions to $\dot{\text{N}}_i=\dot{\text{D}}_i=0$ subject to $\text{N}_i=\text{N}$ and $\text{D}_i=\text{D}$.  With $\text{S}\equiv \text{ND}$ and $\Lambda\equiv\frac{1}{k_c}+\frac{1}{k_t}$, this condition is
\begin{equation}
\beta_\text{N}+\beta_{\text{SN}}\frac{\text{S}^n}{K^n+\text{S}^n}+\frac{\gamma^2}{\Lambda}\frac{\text{S}}{\text{S}-\frac{\beta_\text{D}}{\Lambda}}-\Lambda \text{S} = 0
\end{equation}
We used this equation to determine the homogeneous steady state values $\text{N}^*$ and $\text{D}^*$.

Directly diagonalizing the full Jacobian $J$, a matrix of dimension twice the total number of cells in the lattice, would be very difficult.  Othmer and Scriven \cite{othmer_1971} showed that the problem can be simplified by separating cell adjacency-related intercellular contributions from the intracellular dynamics of the signaling system itself.  This approach first diagonalizes the structure matrix $M$ (in which $M_{ij}=\frac{1}{6}$ where cells $i$ and $j$ are neighbors and $M_{ij}=0$ otherwise) in isolation to yield its spectrum {$q_k$}.  The eigenvalues of $J$ are then the eigenvalues of $H+q_{k}B$, where $H$ and $B$ represent the modulation of production rates due to changes in intra- and extra-cellular components, respectively.  These relationships are represented by the partial derivatives $H_{uv}=\frac{\partial u_{i}}{v_{i}}$ and $B_{uv}=\frac{\partial u_{i}}{\partial v_{j\neq i}}$ where $u$ and $v$ index the chemical species involved in the interaction (here, Notch and Delta) and $i$,$j$ are cell indices.
\begin{equation}
H=\left(\begin{array}{cc} \beta_{\text{SN}}\frac{n}{\text{N}^{*}}f_{0}g_{0}-\gamma-\Lambda\text{D}^{*} & -\frac{1}{k_c}\text{N}^{*} \\ -\frac{1}{k_c}\text{D}^{*} & -\gamma -\Lambda\text{N}^{*} \end{array}\right)
\end{equation}

\begin{equation}
B=\left(\begin{array}{cc} 0 & \beta_{\text{SN}}\frac{n}{\text{D}^{*}}f_{0}g_{0}-\frac{1}{k_t}\text{N}^{*} \\ -\frac{1}{k_t}\text{D}^{*} & 0 \end{array}\right)
\end{equation}
where $f_0\equiv\frac{K_{\text{SN}}^{n}}{K_{\text{SN}}^{n}+(\text{N}^{*}\text{D}^{*})^n}$ and $g_0\equiv\frac{(\text{N}^{*}\text{D}^{*})^n}{K_{\text{SN}}^{n}+(\text{N}^{*}\text{D}^{*})^n}$.  The Othmer and Scriven method thus involves only diagonalizing one large (but sparse) matrix $M$ representing the cell-cell adjacency of the system, and then diagonalizing a small two-by-two matrix.

The characteristic equations of the Jacobian are then
\begin{widetext}
\begin{eqnarray}
&&\left|\begin{array}{cc} \beta_{\text{SN}}\frac{n}{\text{N}^{*}}f_{0}g_{0}-\gamma-\left(\frac{1}{k_t}+\frac{1}{k_c}\right)\text{D}^{*} - \lambda & \left(-\frac{1}{k_c}-q_{k}\frac{1}{k_t}\right)\text{N}^{*} + q_{k}\beta_{\text{SN}}\frac{n}{\text{D}^{*}}f_{0}g_{0} \\ \left(-\frac{1}{k_c}-q_{k}\frac{1}{k_t}\right)\text{D}^{*} & -\gamma -\left(\frac{1}{k_c}+\frac{1}{k_t}\right)\text{N}^{*}-\lambda \end{array}\right| = 0 \nonumber \\
&&\\
&&\rightarrow \lambda=\frac{\frac{\Gamma}{\text{N}}-2\gamma-\Lambda(\text{N}+\text{D})\pm\sqrt{\left(-\frac{\Gamma}{\text{N}}+2\gamma+\Lambda(\text{N}+\text{D})\right)^{2}-4\left(\text{N}\text{D}\left(\Lambda^2-\theta^2\right)+\Lambda\gamma\left(\text{N}+\text{D}\right)+\Gamma\left(q_{k}\theta-\Lambda\right)+\gamma^2-\frac{\Gamma}{\text{N}}\gamma\right)}}{2} \nonumber
\end{eqnarray}
\end{widetext}
where $\Gamma\equiv \beta_{\text{SN}}nf_{0}g_{0}$, $\theta\equiv\frac{1}{k_c}+q_k\frac{1}{k_t}$, and asterisks are omitted.  The sufficient criterion for homogeneous steady state instability is  $\text{N}\text{D}\left(\Lambda^2-\theta^2\right)+\Lambda\gamma\left(\text{N}+\text{D}\right)+\Gamma\left(q_{k}\theta-\Lambda\right)+\gamma^2-\frac{\Gamma}{\text{N}}\gamma<0$.  For the ideal hexagonal lattice of cells with periodic boundary conditions that we consider, the minimum $q_k$ (chosen because it corresponds to the MLE, which is the maximual eigenvalue of $J$) is calculated to be -0.5.

The numerical solutions for the MLE plotted in Fig. \ref{fig:SLIMI}b for particular choices of parameters are illuminating.  Even with no explicitly sharp molecular interactions ($n=1$), there is a sizable region bounded by minimal and maximal $\beta_D$ values over which the MLE is positive, and thus the system may achieve a lateral inhibition pattern as shown in Fig. \ref{fig:SLIMI}b.  A potential shortcoming of the SLIMI mechanism is its sensitivity to leakiness in Notch expression (i.e. non-negligible $\beta_\text{N}$).  Fig. \ref{fig:SLIMI}c plots the MLE profile for progressively greater leakiness in Notch expression indexed as a percentage of maximal Notch induction (\emph{i.e.}, $\beta_{\text{N}}=l\beta_{\text{SN}}$ for $l$ described above the appropriate plot), from which we see that the SLIMI model is fairly robust.

\section{Discussion}
The SLIMI model described here is an extraordinarily simple approach to LI patterning through the Notch-Delta signaling system, conceivable only by virtue of the previously-unappreciated inactivation of Delta by Notch in \emph{cis}.  It is appealing for a number of reasons.  First, we have shown that SLIMI is a feedback architecture that supports the formation of the LI pattern across a wide range of parameters.  Second, it does so even in the presence of non-ideal leakiness in the regulatory feedback.  Third, and perhaps most dramatically, it does not require the action of any hypothetical intermediate factor.  The sole necessary regulatory interaction has been shown to occur naturally \cite{huppert_1997, wilkinson_1994} in at least some contexts.

Appealing though these features of SLIMI may be, it remains uncertain if natural systems in fact utilize this feedback to generate LI patterns.  Results regarding Notch-Delta feedback elements operating in lateral inhibition patterning processes are varied at present, seemingly indicating a degree of context specificity \cite{seugnet_1997, parks_2008} that defies efforts to postulate a universal mechanism.  As mentioned earlier, there is some evidence for increased Notch expression rates induced by Notch signaling in LI-patterning systems \cite{wilkinson_1994}, but the contribution of this relative to that of other feedbacks in driving the patterning process is unknown.


\begin{thebibliography}{18}%
\makeatletter
\providecommand \@ifxundefined [1]{%
 \@ifx{#1\undefined}
}%
\providecommand \@ifnum [1]{%
 \ifnum #1\expandafter \@firstoftwo
 \else \expandafter \@secondoftwo
 \fi
}%
\providecommand \@ifx [1]{%
 \ifx #1\expandafter \@firstoftwo
 \else \expandafter \@secondoftwo
 \fi
}%
\providecommand \natexlab [1]{#1}%
\providecommand \enquote  [1]{``#1''}%
\providecommand \bibnamefont  [1]{#1}%
\providecommand \bibfnamefont [1]{#1}%
\providecommand \citenamefont [1]{#1}%
\providecommand \href@noop [0]{\@secondoftwo}%
\providecommand \href [0]{\begingroup \@sanitize@url \@href}%
\providecommand \@href[1]{\@@startlink{#1}\@@href}%
\providecommand \@@href[1]{\endgroup#1\@@endlink}%
\providecommand \@sanitize@url [0]{\catcode `\\12\catcode `\$12\catcode
  `\&12\catcode `\#12\catcode `\^12\catcode `\_12\catcode `\%12\relax}%
\providecommand \@@startlink[1]{}%
\providecommand \@@endlink[0]{}%
\providecommand \url  [0]{\begingroup\@sanitize@url \@url }%
\providecommand \@url [1]{\endgroup\@href {#1}{\urlprefix }}%
\providecommand \urlprefix  [0]{URL }%
\providecommand \Eprint [0]{\href }%
\@ifxundefined \urlstyle {%
  \providecommand \doi  [0]{\begingroup \@sanitize@url \@doi}%
  \providecommand \@doi [1]{\endgroup \@@startlink {\doibase
  #1}doi:\discretionary {}{}{}#1\@@endlink }%
}{%
  \providecommand \doi  [0]{doi:\discretionary{}{}{}\begingroup
  \urlstyle{rm}\Url }%
}%
\providecommand \doibase [0]{http://dx.doi.org/}%
\providecommand \Doi [0]{\begingroup \@sanitize@url \@Doi }%
\providecommand \@Doi  [1]{\endgroup\@@startlink{\doibase#1}\@@Doi}%
\providecommand \@@Doi [1]{#1\@@endlink}%
\providecommand \selectlanguage [0]{\@gobble}%
\providecommand \bibinfo  [0]{\@secondoftwo}%
\providecommand \bibfield  [0]{\@secondoftwo}%
\providecommand \translation [1]{[#1]}%
\providecommand \BibitemOpen [0]{}%
\providecommand \bibitemStop [0]{}%
\providecommand \bibitemNoStop [0]{.\EOS\space}%
\providecommand \EOS [0]{\spacefactor3000\relax}%
\providecommand \BibitemShut  [1]{\csname bibitem#1\endcsname}%
\bibitem [{\citenamefont {Reed}(2004)}]{reed_2004}%
  \BibitemOpen
  \bibfield  {author} {\bibinfo {author} {\bibfnamefont {R.~D.}\ \bibnamefont
  {Reed}},\ }\href@noop {} {\bibfield  {journal} {\bibinfo  {journal}
  {Development Genes and Evolution},\ }\textbf {\bibinfo {volume} {214}},\
  \bibinfo {pages} {43} (\bibinfo {year} {2004})}\BibitemShut {NoStop}%
\bibitem [{\citenamefont {Kunisch}\ \emph {et~al.}(1994)\citenamefont
  {Kunisch}, \citenamefont {Haenlin},\ and\ \citenamefont
  {Campos-Ortega}}]{campos-ortega_1994}%
  \BibitemOpen
  \bibfield  {author} {\bibinfo {author} {\bibfnamefont {M.}~\bibnamefont
  {Kunisch}}, \bibinfo {author} {\bibfnamefont {M.}~\bibnamefont {Haenlin}}, \
  and\ \bibinfo {author} {\bibfnamefont {J.~A.}\ \bibnamefont
  {Campos-Ortega}},\ }\href@noop {} {\bibfield  {journal} {\bibinfo  {journal}
  {Proceedings of the National Academy of Science USA},\ }\textbf {\bibinfo
  {volume} {91}},\ \bibinfo {pages} {10139} (\bibinfo {year}
  {1994})}\BibitemShut {NoStop}%
\bibitem [{\citenamefont {Marnellos}\ \emph {et~al.}(2000)\citenamefont
  {Marnellos}, \citenamefont {Deblandre}, \citenamefont {Mjolsness},\ and\
  \citenamefont {Kinter}}]{marnellos_2000}%
  \BibitemOpen
  \bibfield  {author} {\bibinfo {author} {\bibfnamefont {G.}~\bibnamefont
  {Marnellos}}, \bibinfo {author} {\bibfnamefont {G.}~\bibnamefont
  {Deblandre}}, \bibinfo {author} {\bibfnamefont {E.}~\bibnamefont
  {Mjolsness}}, \ and\ \bibinfo {author} {\bibfnamefont {C.}~\bibnamefont
  {Kinter}},\ }\href@noop {} {\bibfield  {journal} {\bibinfo  {journal}
  {Pacific Symposium on Biocomputing},\ }\textbf {\bibinfo {volume} {5}},\
  \bibinfo {pages} {326} (\bibinfo {year} {2000})}\BibitemShut {NoStop}%
\bibitem [{\citenamefont {Heitzler}\ and\ \citenamefont
  {Simpson}(1991)}]{heitzler_1991}%
  \BibitemOpen
  \bibfield  {author} {\bibinfo {author} {\bibfnamefont {P.}~\bibnamefont
  {Heitzler}}\ and\ \bibinfo {author} {\bibfnamefont {P.}~\bibnamefont
  {Simpson}},\ }\href@noop {} {\bibfield  {journal} {\bibinfo  {journal}
  {Cell},\ }\textbf {\bibinfo {volume} {64}},\ \bibinfo {pages} {1083}
  (\bibinfo {year} {1991})}\BibitemShut {NoStop}%
\bibitem [{\citenamefont {D'Souza}\ \emph {et~al.}(2008)\citenamefont
  {D'Souza}, \citenamefont {Miyamoto},\ and\ \citenamefont
  {Weinmaster}}]{weinmaster_2008}%
  \BibitemOpen
  \bibfield  {author} {\bibinfo {author} {\bibfnamefont {B.}~\bibnamefont
  {D'Souza}}, \bibinfo {author} {\bibfnamefont {A.}~\bibnamefont {Miyamoto}}, \
  and\ \bibinfo {author} {\bibfnamefont {G.}~\bibnamefont {Weinmaster}},\
  }\href@noop {} {\bibfield  {journal} {\bibinfo  {journal} {Oncogene},\
  }\textbf {\bibinfo {volume} {27}},\ \bibinfo {pages} {5148} (\bibinfo {year}
  {2008})}\BibitemShut {NoStop}%
\bibitem [{\citenamefont {Weinmaster}\ and\ \citenamefont
  {Kopan}(2006)}]{weinmaster_kopan_2006}%
  \BibitemOpen
  \bibfield  {author} {\bibinfo {author} {\bibfnamefont {G.}~\bibnamefont
  {Weinmaster}}\ and\ \bibinfo {author} {\bibfnamefont {R.}~\bibnamefont
  {Kopan}},\ }\href@noop {} {\bibfield  {journal} {\bibinfo  {journal}
  {Development},\ }\textbf {\bibinfo {volume} {133}},\ \bibinfo {pages} {3277}
  (\bibinfo {year} {2006})}\BibitemShut {NoStop}%
\bibitem [{\citenamefont {Bray}(2006)}]{bray_2006}%
  \BibitemOpen
  \bibfield  {author} {\bibinfo {author} {\bibfnamefont {S.~J.}\ \bibnamefont
  {Bray}},\ }\href@noop {} {\bibfield  {journal} {\bibinfo  {journal} {Nature
  Reviews Molecular Cell Biology},\ }\textbf {\bibinfo {volume} {7}},\ \bibinfo
  {pages} {678} (\bibinfo {year} {2006})}\BibitemShut {NoStop}%
\bibitem [{\citenamefont {Artavanis-Tsakonas}\ \emph
  {et~al.}(1999)\citenamefont {Artavanis-Tsakonas}, \citenamefont {Rand},\ and\
  \citenamefont {Lake}}]{artavanis-tsakonas_1999}%
  \BibitemOpen
  \bibfield  {author} {\bibinfo {author} {\bibfnamefont {S.}~\bibnamefont
  {Artavanis-Tsakonas}}, \bibinfo {author} {\bibfnamefont {M.~D.}\ \bibnamefont
  {Rand}}, \ and\ \bibinfo {author} {\bibfnamefont {R.~J.}\ \bibnamefont
  {Lake}},\ }\href@noop {} {\bibfield  {journal} {\bibinfo  {journal}
  {Science},\ }\textbf {\bibinfo {volume} {284}},\ \bibinfo {pages} {770}
  (\bibinfo {year} {1999})}\BibitemShut {NoStop}%
\bibitem [{\citenamefont {Collier}\ \emph {et~al.}(1996)\citenamefont
  {Collier}, \citenamefont {Monk}, \citenamefont {Maini},\ and\ \citenamefont
  {Lewis}}]{lewis_1996}%
  \BibitemOpen
  \bibfield  {author} {\bibinfo {author} {\bibfnamefont {J.~R.}\ \bibnamefont
  {Collier}}, \bibinfo {author} {\bibfnamefont {N.~A.}\ \bibnamefont {Monk}},
  \bibinfo {author} {\bibfnamefont {P.~K.}\ \bibnamefont {Maini}}, \ and\
  \bibinfo {author} {\bibfnamefont {J.~H.}\ \bibnamefont {Lewis}},\ }\href@noop
  {} {\bibfield  {journal} {\bibinfo  {journal} {Journal of Theoretical
  Biology},\ }\textbf {\bibinfo {volume} {183}},\ \bibinfo {pages} {429}
  (\bibinfo {year} {1996})}\BibitemShut {NoStop}%
\bibitem [{\citenamefont {Plahte}\ and\ \citenamefont
  {{\O}yehaug}(2007)}]{plahte_2007}%
  \BibitemOpen
  \bibfield  {author} {\bibinfo {author} {\bibfnamefont {E.}~\bibnamefont
  {Plahte}}\ and\ \bibinfo {author} {\bibfnamefont {L.}~\bibnamefont
  {{\O}yehaug}},\ }\href@noop {} {\bibfield  {journal} {\bibinfo  {journal}
  {Physica D: Nonlinear Phenomena},\ }\textbf {\bibinfo {volume} {226}},\
  \bibinfo {pages} {117} (\bibinfo {year} {2007})}\BibitemShut {NoStop}%
\bibitem [{\citenamefont {Lecourtois}\ and\ \citenamefont
  {Schweisguth}(1995)}]{lecourtois_1995}%
  \BibitemOpen
  \bibfield  {author} {\bibinfo {author} {\bibfnamefont {M.}~\bibnamefont
  {Lecourtois}}\ and\ \bibinfo {author} {\bibfnamefont {F.}~\bibnamefont
  {Schweisguth}},\ }\href@noop {} {\bibfield  {journal} {\bibinfo  {journal}
  {Genes \& Development},\ }\textbf {\bibinfo {volume} {9}},\ \bibinfo {pages}
  {2598} (\bibinfo {year} {1995})}\BibitemShut {NoStop}%
\bibitem [{\citenamefont {Greenwald}(1998)}]{greenwald_1998}%
  \BibitemOpen
  \bibfield  {author} {\bibinfo {author} {\bibfnamefont {I.}~\bibnamefont
  {Greenwald}},\ }\href@noop {} {\bibfield  {journal} {\bibinfo  {journal}
  {Genes \& Development},\ }\textbf {\bibinfo {volume} {12}},\ \bibinfo {pages}
  {1751} (\bibinfo {year} {1998})}\BibitemShut {NoStop}%
\bibitem [{\citenamefont {Huppert}\ \emph {et~al.}(1997)\citenamefont
  {Huppert}, \citenamefont {Jacobsen},\ and\ \citenamefont
  {Muskavitch}}]{huppert_1997}%
  \BibitemOpen
  \bibfield  {author} {\bibinfo {author} {\bibfnamefont {S.~S.}\ \bibnamefont
  {Huppert}}, \bibinfo {author} {\bibfnamefont {T.~L.}\ \bibnamefont
  {Jacobsen}}, \ and\ \bibinfo {author} {\bibfnamefont {M.~A.}\ \bibnamefont
  {Muskavitch}},\ }\href@noop {} {\bibfield  {journal} {\bibinfo  {journal}
  {Development},\ }\textbf {\bibinfo {volume} {124}},\ \bibinfo {pages} {3283}
  (\bibinfo {year} {1997})}\BibitemShut {NoStop}%
\bibitem [{\citenamefont {Wilkinson}\ \emph {et~al.}(1994)\citenamefont
  {Wilkinson}, \citenamefont {Fitzgerald},\ and\ \citenamefont
  {Greenwald}}]{wilkinson_1994}%
  \BibitemOpen
  \bibfield  {author} {\bibinfo {author} {\bibfnamefont {H.~A.}\ \bibnamefont
  {Wilkinson}}, \bibinfo {author} {\bibfnamefont {K.}~\bibnamefont
  {Fitzgerald}}, \ and\ \bibinfo {author} {\bibfnamefont {I.}~\bibnamefont
  {Greenwald}},\ }\href@noop {} {\bibfield  {journal} {\bibinfo  {journal}
  {Cell},\ }\textbf {\bibinfo {volume} {79}},\ \bibinfo {pages} {1187}
  (\bibinfo {year} {1994})}\BibitemShut {NoStop}%
\bibitem [{\citenamefont {Sprinzak}\ \emph {et~al.}(2010)\citenamefont
  {Sprinzak}, \citenamefont {Lakhanpal}, \citenamefont {LeBon}, \citenamefont
  {Santat}, \citenamefont {Fontes}, \citenamefont {Anderson}, \citenamefont
  {Garcia-Ojalvo},\ and\ \citenamefont {Elowitz}}]{sprinzak_2010}%
  \BibitemOpen
  \bibfield  {author} {\bibinfo {author} {\bibfnamefont {D.}~\bibnamefont
  {Sprinzak}}, \bibinfo {author} {\bibfnamefont {A.}~\bibnamefont {Lakhanpal}},
  \bibinfo {author} {\bibfnamefont {L.}~\bibnamefont {LeBon}}, \bibinfo
  {author} {\bibfnamefont {L.~A.}\ \bibnamefont {Santat}}, \bibinfo {author}
  {\bibfnamefont {M.~E.}\ \bibnamefont {Fontes}}, \bibinfo {author}
  {\bibfnamefont {G.~A.}\ \bibnamefont {Anderson}}, \bibinfo {author}
  {\bibfnamefont {J.}~\bibnamefont {Garcia-Ojalvo}}, \ and\ \bibinfo {author}
  {\bibfnamefont {M.~B.}\ \bibnamefont {Elowitz}},\ }\href@noop {} {\bibfield
  {journal} {\bibinfo  {journal} {Nature},\ }\textbf {\bibinfo {volume}
  {465}},\ \bibinfo {pages} {86} (\bibinfo {year} {2010})}\BibitemShut
  {NoStop}%
\bibitem [{\citenamefont {Othmer}\ and\ \citenamefont
  {Scriven}(1971)}]{othmer_1971}%
  \BibitemOpen
  \bibfield  {author} {\bibinfo {author} {\bibfnamefont {H.}~\bibnamefont
  {Othmer}}\ and\ \bibinfo {author} {\bibfnamefont {L.}~\bibnamefont
  {Scriven}},\ }\href@noop {} {\bibfield  {journal} {\bibinfo  {journal}
  {Journal of Theoretical Biology},\ }\textbf {\bibinfo {volume} {32}},\
  \bibinfo {pages} {507} (\bibinfo {year} {1971})}\BibitemShut {NoStop}%
\bibitem [{\citenamefont {Seugnet}\ \emph {et~al.}(1997)\citenamefont
  {Seugnet}, \citenamefont {Simpson},\ and\ \citenamefont
  {Haenlin}}]{seugnet_1997}%
  \BibitemOpen
  \bibfield  {author} {\bibinfo {author} {\bibfnamefont {L.}~\bibnamefont
  {Seugnet}}, \bibinfo {author} {\bibfnamefont {P.}~\bibnamefont {Simpson}}, \
  and\ \bibinfo {author} {\bibfnamefont {M.}~\bibnamefont {Haenlin}},\
  }\href@noop {} {\bibfield  {journal} {\bibinfo  {journal} {Development},\
  }\textbf {\bibinfo {volume} {124}},\ \bibinfo {pages} {2015} (\bibinfo {year}
  {1997})}\BibitemShut {NoStop}%
\bibitem [{\citenamefont {Parks}\ \emph {et~al.}(2008)\citenamefont {Parks},
  \citenamefont {Shalaby},\ and\ \citenamefont {Muskavitch}}]{parks_2008}%
  \BibitemOpen
  \bibfield  {author} {\bibinfo {author} {\bibfnamefont {A.~L.}\ \bibnamefont
  {Parks}}, \bibinfo {author} {\bibfnamefont {N.~A.}\ \bibnamefont {Shalaby}},
  \ and\ \bibinfo {author} {\bibfnamefont {M.~A.}\ \bibnamefont {Muskavitch}},\
  }\href@noop {} {\bibfield  {journal} {\bibinfo  {journal} {Genesis},\
  }\textbf {\bibinfo {volume} {46}},\ \bibinfo {pages} {265} (\bibinfo {year}
  {2008})}\BibitemShut {NoStop}%
\end{thebibliography}
\end{document}